\documentclass[]{interact}

\usepackage{epstopdf}
\usepackage[caption=false]{subfig}
\usepackage[numbers,sort&compress]{natbib}
\bibpunct[, ]{[}{]}{,}{n}{,}{,}

\theoremstyle{plain}

\theoremstyle{definition}

\theoremstyle{remark}

\newcommand{\iom}{i\omega_n}
\newcommand{\vek}{\vec{k}}
\newcommand{\veq}{\vec{k}+\vec{Q}}
\usepackage{amsmath}
\usepackage{float}

\usepackage[T1]{fontenc}
\begin{document}


\title{Coexistence of superconductivity and charge density wave in a correlated regime}

\author{
\name{E. J. Calegari\textsuperscript{a}\thanks{CONTACT E. J. Calegari. Email: eleonir@ufsm.br}, L. C. Prauchner\textsuperscript{b}, A. C. Lausmann\textsuperscript{a} and S. G. Magalhaes\textsuperscript{b}}
\affil{\textsuperscript{a}Departamento de F\'{\i}sica, Universidade Federal de Santa Maria, 97105-900, Santa Maria, RS, Brazil; \textsuperscript{b} Instituto de Física, Universidade Federal do Rio Grande do Sul, 91501-970, Porto Alegre, RS, Brazil
}}

\maketitle

\begin{abstract}
To investigate the coexistence of
superconductivity and charge density wave (CDW) in a correlated regime, we  employ the Green's functions formalism, as well as the Hubbard-I approximation, as a way to introduce the correlations into the problem, in the form of a repulsive Coulomb interaction $U$. In addition, we investigate the effects of second-nearest neighbor hopping $t_1$ on a pure CDW state. The analysis of the  results show that, for small values of $t_1$, both CDW and superconducting gaps compete for the same region on the Fermi surface. The increase of $t_1$ decreases the competition and may lead the system to a coexistence regime. Effects of temperature in the coexistence regime, are also investigated.

\end{abstract}


\section{Introduction}

Over the past decades, strongly correlated electron systems  have produced a wealth of intriguing experimental results that continue to pose significant challenges for theory. Notable examples include the metal-insulator transition \cite{Imada1998}, the competition or coexistence between Kondo effect and magnetism \cite{Coqblin2008, Coqblin2006, Magalhaes2006}, heavy-fermion superconductors \cite{Steglich1991, Cox1995} and cuprate superconductors \cite{Taillefer2010, Proust2018}.
This last category of systems has a rich phase diagram, with many phases and coexistence regimes.
Interestingly, it is also an example in which both charge density wave (CDW) and superconductivity (SC) coexist \cite{hayden}, while also presenting the pseudogap phenomenon, antiferromagnetism, and strange metal behavior \cite{Proust2018}. There is a consensus that the mechanisms that generate superconductivity in these systems are directly associated with the other neighboring phases.
Another category  of strongly correlated materials in which SC and CDW coexist is the transition metal dichalcogenides (TMDs) \cite{chen,morosan,lian,Venditti}.  Recently, superconductivity was discovered in nickelate materials \cite{Lee}, which are also strongly correlated systems. An interesting result was reported for the nickelate La$_4$Ni$_3$O$_{10}$, which at low pressure exhibits an intricate interplay between concurrent spin density wave (SDW) and CDW. However, as pressure increases, the density wave orders are suppressed and superconductivity suddenly emerges, signaling robust competition between density wave orders and superconductivity \cite{wang}.

SC and CDW share differences and similarities \cite{xu} and are collective phenomena that are usually adjacent in phase diagrams. In general, the state of coexistence can be tuned using a control parameter such as pressure or doping \cite{morosan,Venditti,morosan2}.
 In general, with the increase of external pressure, the lattice parameters decrease \cite{Jiang}, causing an increase not only in  first-nearest neighbor hopping $t_0$ \cite{sil}, but also in second-nearest neighbor hopping $t_1$ \cite{araujo}. The presence of $t_1$ can shift the band energy in the region of antinodal point $(\pi,0)$ destroying the perfect nesting and suppressing the CDW correlations \cite{Veric}. For this reason, in the present work we focused on the second-nearest neighbor hopping, which
plays a relevant role in the investigation of the CDW instability and its coexistence with superconductivity.

It is important to emphasize that the presence of correlations in the aforementioned systems makes their  investigation a complex task. As a consequence, the mechanisms behind the coexistence of SC and CDW remain an open question in condensed matter physics.

With the aim of shedding light on this issue, in this work
we analyze the competition, or possible coexistence, between $s$-wave superconductivity and charge density wave phases under the effect of local repulsive Coulomb interaction $U$. The second-nearest neighbor hopping amplitude $t_1$ is used as a control parameter, varied to tune the coexistence region. Furthermore, we investigate the effects of correlations
on the CDW phase as well as on the coexistence of SC and CDW.
The methodology considered here \cite{jesus350, calegari2016,sampaio,prauchner} allows us to obtain Green's functions in which both instabilities are treated within the BCS framework \cite{BCS,balseiro}. To account for correlations,
the normal-state uncorrelated Green’s functions are replaced by those obtained via Hubbard-I approximation applied to the
one-band Hubbard model \cite{hubbard}, which is traditionally utilized to study strongly correlated systems.
This methodology enables us to investigate the interplay between SC and CDW in the presence of correlations.  Since we use neither a specific dispersion relation nor material-realistic hopping parameters, we do not intend to describe any specific system with the model proposed in the present work. However, the present model and methodology provide valuable insights into relevant systems, as for instance, the featuring TMD compound 1$T$-TaS$_2$ \cite{luca,dalal} and the nickelate Nd$_{1-x}$Sr$_x$NiO$_2$ \cite{Motoharu}.

This article is organized as follows: Section 2 presents the theory and model. In Section 3, we show the main numerical results, including the phase diagrams and  spectral function. Finally, Section 4 discusses the main results and present the conclusions.

\section{The Model}
We start with a model whose Hamiltonian reads:
\begin{equation}
     \mathcal{\hat{H}} =\sum_{\vec{k},\sigma} \xi_{\vec{k}} \ \hat{c}_{\vec{k},\sigma}^{\dagger} \hat{c}_{\vec{k},\sigma}+\hat{H}_{PAR}+\hat{H}_{CDW},
      \label{H1}
\end{equation}
where $\xi_{\vec{k}} = \varepsilon_{\vec{k}} - \mu$, with $\varepsilon_{\vec{k}}$ being the dispersion relation for a square lattice and $\mu$ the chemical potential.
The pairing term \cite{tinkham} is given by:
\begin{equation}
\hat{H}_{PAR}=\sum_{\vec{k},{\vec{k}}^{'}}V_{\vec{k},{\vec{k}}^{'}}\hat{c}^{\dagger}_{\vec{k}, {\uparrow}} \hat{c}^{\dagger}_{-\vec{k},{\downarrow}}\hat{c}_{-\vec{k}^{'},{\downarrow}} \hat{c}_{\vec{k}^{'},{\uparrow}}
\label{Hbcs}
\end{equation}
while the term responsible for the formation of a CDW \cite{prauchner,balseiro} is
\begin{equation}
     \hat{H}_{CDW} = \sum_{\vec{k},\vec{k'},\sigma,\sigma'} V_{\vec{k}\vec{k'}} \hat{c}_{{\vec{k}+\vec{Q}},\sigma}^{\dagger}\hat{c}_{\vec{k},\sigma}\hat{c}_{{\vec{k'}+\vec{Q}},\sigma'}^{\dagger}\hat{c}_{\vec{k'},\sigma'}.
      \label{Hcdw}
\end{equation}
$\hat{c}_{k,\sigma}^{\dagger}$ is the fermionic creation operator while  $\hat{c}_{k,\sigma}$ is the fermionic annihilation operator and $\sigma(=\uparrow,\downarrow)$ is the spin index.
$\vec{Q}$ is the nesting vector which satisfies the condition $\vec{k}+2\vec{Q}=\vec{k}$. The terms $\hat{H}_{PAR}$ and $\hat{H}_{CDW}$ are treated in the BCS level \cite{BCS,balseiro} and the origin of the attractive interaction $V_{\vec{k}\vec{k'}}$ is left unspecified \cite{jesus350,sampaio}.
When treated in the mean field level, the Hamiltonian given in Eq.(\ref{H1}) becomes:
\begin{equation}
\label{HMF}
    \begin{split}
    \mathcal{\hat{H}}_{MF} =& \sum_{\vec{k},\sigma}  \xi_{\vec{k}} \hat{c}_{\vec{k},\sigma}^{\dagger} \hat{c}_{\vec{k},\sigma}
       +\sum_{\vec{k}}\big(\Delta_{\vec{k}}^* \hat{c}_{\vec{k},\uparrow}\hat{c}_{-\vec{k},\downarrow} +\Delta_{\vec{k}} \hat{c}_{-\vec{k},\downarrow}^{\dagger} \hat{c}_{\vec{k},\uparrow}^{\dagger}  \\&+    W_{\vec{k}}\hat{c}_{\vec{k}+\vec{Q},\uparrow}^{\dagger}\hat{c}_{\vec{k},\uparrow} + W_{\vec{k}} \hat{c}_{\vec{k}+\vec{Q},\downarrow}^{\dagger}\hat{c}_{\vec{k},\downarrow} \big),
    \end{split}
\end{equation}
where $\Delta_{\vec{k}}$ and $W_{\vec{k}}$ are respectively, the superconducting and the CDW order parameters, which are defined as:
\begin{equation}
   \Delta_{\vec{k}} = \sum_{\vec{k'}}V_{\vec{k}\vec{k'}}\langle \hat{c}_{-\vec{k'},\downarrow}\hat{c}_{\vec{k'},\uparrow} \rangle
\label{Delta}
\end{equation}
and
\begin{equation}
 W_{\vec{k}} = \sum_{\vec{k'}}V_{\vec{k}\vec{k'}}\langle \hat{c}_{\vec{k'}+\vec{Q},\uparrow}^{\dagger}\hat{c}_{\vec{k'},\uparrow} \rangle.
 \end{equation}
 At this point, we have deviated somewhat from the work in reference \cite{balseiro}, which considers the description of the CDW phase in more detail and takes into account two order parameters, $G_0$ and $G_1$. Basically, the parameter $G_0$ results from a sum over $\vec{k}$ considering selected states that satisfy both conditions, $|\varepsilon_{\vec{k}} - \varepsilon_F|<\varepsilon_{CDW}$ and $|\varepsilon_{\vec{k}+\vec{Q}} - \varepsilon_F|<\varepsilon_{CDW}$, in which $\varepsilon_{CDW}$ and $\varepsilon_F$ are the cutoff and Fermi energies, respectively. The other parameter, $G_1$, results from a sum over $\vec{k}$ without restrictions.  Assuming that the most important states for the CDW are those closest to the Fermi surface, in the present work only the order parameter equivalent to $G_0$  was considered. It should be emphasized that this choice yields results that are, at least qualitatively, in agreement with the findings of Balseiro and Falicov \cite{balseiro}.

Following the BCS framework, we replace the interaction $V_{\vec{k}\vec{k'}}$ by the simpler form

\begin{equation}
V_{\vec{k}\vec{k'}}=
\begin{cases}
-V,\ \text{if} \ |\varepsilon_{\vec{k}} - \mu| < \varepsilon_c \\
-V,\ \text{if} \ |\varepsilon_{\vec{k}+\vec{Q}} - \mu| < \varepsilon_c  \\
0, \ \text{otherwise},
\end{cases}
\label{Vkk}
\end{equation}
%
with $\varepsilon_c=\varepsilon_{SC}$ for $\Delta_{\vec{k}}$, and $\varepsilon_c=\varepsilon_{CDW}$, for $W_{\vec{k}}$. The dispersion relation for a square lattice is given by:
\begin{equation}
        \varepsilon_{\vec{k}} =2t_0[\cos(k_xa)+\cos(k_ya)] + 4t_1\cos(k_xa)\cos(k_ya),
    \end{equation}
with $t_0$ representing the first-nearest neighbor hopping and $t_1$ the second-nearest neighbor hopping while $a$ is the lattice constant.

To derive the mathematical expressions for the order parameters $\Delta_{\vec{k}}$ and $W_{\vec{k}}$, we obtain the following result
\begin{equation}
    (i\omega_n{\bf{I}} - H_{\vec{k}})\mathcal{G}(\vec{k},i\omega_n)  = {\bf{I}},
   \label{eqGI}
\end{equation}
where $\mathcal{G}(\vec{k},i\omega_n)$ is the Green's function matrix for the mean field Hamiltonian (see Eq. (\ref{HMF})), $\bf{I}$ is a fourth-order identity matrix and the coefficients matrix $H_{\vec{k}}$ is given by:
$$
H_{\vec{k}} =
\begin{pmatrix}\label{eq:matrizCoeff}
\xi_{\vec{k}} & -\Delta_{\vek} & W_{\vek} & 0 \\
-\Delta_{\vek}^* & -\xi_{\vek} & 0 & -W_{\vek} \\
W_{\vek} & 0 & \xi_{\veq} & -\Delta_{\veq} \\
0 & -W_{\vek} & -\Delta_{\veq}^* & -\xi_{\veq}
\end{pmatrix}.
$$

As long as $(i\omega_n{\bf{I}} - H_{\vec{k}})$ is not a null matrix, we can solve Eq. (\ref{eqGI}), obtaining
\begin{equation}
    \mathcal{G}(\vec{k},i\omega_n) = (i\omega_n{\bf{I}} - H_{\vec{k}})^{-1}{\bf{I}}.
   \label{eqGI1}
\end{equation}

In the normal state $W_{\vek}=0$, as well as $\Delta_{\vek}=\Delta_{\veq}=0$. In this case, only the elements of the main diagonal of the Green's function matrix $\mathcal{G}(\vec{k},i\omega_n)$, remain non-null. These elements are:
\begin{align}
\label{G0up}
 G_{0\uparrow}(\vek,\iom)&=(\iom-\xi_{\vek})^{-1},\\
 G_{0\downarrow}(\vek,\iom)&=-(\iom+\xi_{\vek})^{-1},\\
 G_{0\uparrow}(\veq,\iom)&=(\iom-\xi_{\veq})^{-1}
 \end{align}
and
\begin{equation}
  G_{0\downarrow}(\veq,\iom)=-(\iom+\xi_{\veq})^{-1}.
  \label{G0down}
\end{equation}
In terms of Green's functions given in Eqs. (\ref{G0up})-(\ref{G0down}), the quantity $i\omega_n{\bf{I}} - H_{\vec{k}}$ present in Eq. (\ref{eqGI1}) can be rewritten as:
$$
i\omega_n{\bf{I}} - H_{\vec{k}} =
\begin{pmatrix}
G_{0\uparrow}^{-1}(\vek) & \Delta_{\vek} & -W_{\vek} & 0 \\
-\Delta_{\vek}^* & G_{0\downarrow}^{-1}(\vek) & 0 & -W_{\vek} \\
-W_{\vec{k}} & 0 & G_{0\uparrow}^{-1}(\vec{K}) & \Delta_{\vec{K}}  \\
0 & -W_{\vec{k}} & -\Delta_{\vec{K}}^* & G_{0\downarrow}^{-1}(\vec{K})
\end{pmatrix},
$$
where $\vec{K} = \veq$ and for simplicity, the dependence of $G_0$ on $\iom$ is omitted.
By using the result above in Eq. (\ref{eqGI1}), we get the elements of the Green's function matrix in the following form:
\begin{equation}
    G^{(nm)}(\vek,\iom)=\frac{A^{(nm)}_{\vek}(\iom)}{D_{\vek}(\iom)},
\end{equation}
where $D_{\vek}(\iom)$ is the determinant of the matrix $i\omega_n{\bf{I}} - H_{\vec{k}}$,
\begin{equation}
 \begin{split}
    D_{\vek}(\iom)&= [G^{-1}_{0\downarrow}(\vek)G^{-1}_{0\downarrow}(\vec{K})-W^{2}_{\vek}][G^{-1}_{0\uparrow}(\vek)G^{-1}_{0\uparrow}(\vec{K})-W^{2}_{\vek}]\\
    &+\Delta^2_{\vec{k}}[G^{-1}_{0\downarrow}(\vec{K})G^{-1}_{0\uparrow}(\vec{K})+\Delta^2_{\vec{K}}]+2\Delta_{\vec{k}}\Delta_{\vec{K}}W^{2}_{\vek}\\
    &
    +\Delta^2_{\vec{K}}G^{-1}_{0\downarrow}(\vec{k})G^{-1}_{0\uparrow}(\vec{k}).
    \end{split}
    \label{Dw}
\end{equation}

In the present work, the most important elements of the Green's function matrix $\mathcal{G}(\vec{k},i\omega_n)$ are $G^{(11)}$, $G^{(22)}$, $G^{(33)}$, $G^{(44)}$,  which allow us to obtain the total occupation number $n_{T}$. Furthermore,  $G^{(12)}$ and $G^{(13)}$ allow for the calculation of the SC
and CDW order parameters, respectively.
Defining:
\begin{equation}
    A_{\sigma,\vek} = B_{\vec{K}}G^{-1}_{0,-\sigma}(\vek) -G^{-1}_{0,\sigma}(\vec{K})W^2_{\vek},
\end{equation}
the numerators $A^{(nm)}$ for the Green's functions are:
\begin{align}
   A^{(11)}_{\vek}&=A_{\uparrow,\vek},\\
A^{(22)}_{\vek}& =  A_{\downarrow,\vek},\\
   A^{(21)}_{\vek}&= -\Delta_{\vec{K}}W^2_{\vek} - B_{\vec{K}}\Delta_{\vek},\\
  A^{(13)}_{\vek}&= -W_{\vek}[\Delta_{\vek}\Delta_{\vec{K}}-G_{0,\downarrow}^{-1}(\vec{k})G_{0,\downarrow}^{-1}(\vec{K}) +W^2_{\vek}],
  \label{Anm}
 \end{align}
with $B_{\vec{K}} =G^{-1}_{0\downarrow}(\vec{K})G^{-1}_{0\uparrow}(\vec{K}) + \Delta^2_{\vec{K}}$.
Interchanging $\vec{k}$ and $\vec{K}$in  $A^{(11)}_{\vek}$, we obtain $A^{(33)}_{\vec{K}}$. Repeating the same process in $A^{(22)}_{\vek}$, we get $A^{(44)}_{\vec{K}}$.


In order to introduce the correlations associated with the Coulomb interaction, the Green's functions with subscript $0$ are replaced by their correlated versions obtained from the one-band Hubbard model \cite{hubbard}, whose Hamiltonian is given by:
\begin{equation}\label{eq:modelohubbard}
     \hat{H} = \sum_{\langle \langle i,j\rangle \rangle,\sigma} t_{i,j} \hat{c}_{i, \sigma}^{\dagger}\hat{c}_{j, \sigma} + \frac{U}{2}\sum_{i,\sigma}\hat{n}_{i, \sigma}\hat{n}_{i, -\sigma}.
 \end{equation}
In the first term of $\hat{H}$, the symbol $\langle\langle...\rangle\rangle$ indicates a sum over
ﬁrst- and second-nearest neighbors while $\hat{c}_{i,\sigma}^{\dagger}$ is the fermionic creation operator and $\hat{c}_{j,\sigma}$ is the fermionic annihilation operator. The second term in $\hat{H}$ considers the repulsive Coulomb interaction $U$ between electrons on the same site $i$.
Following the Hubbard-I approximation \cite{hubbard} for the Hubbard model, we obtain the correlated Green's function for the normal state:
\begin{equation}
G_{\sigma,N}(\vec{k},i\omega_n)= \frac{i\omega_n-U(1-n_{-\sigma})}{(i\omega_n -\varepsilon_{\vec{k}})(i\omega_n-U)-n_{-\sigma}U\varepsilon_{\vec{k}}}.
\label{eqGN}
\end{equation}

To investigate the effects of the Coulomb interaction on  SC and CDW, at this point we replace the Green's functions $G_{0,\sigma}$ present in Eqs. (\ref{Dw})-(\ref{Anm}), by their correlated version $G_{\sigma,N}$, given in Eq. (\ref{eqGN}). It is assumed that the Coulomb interaction  does not significantly affect the BCS formalism \cite{jesus350,calegari2016,sampaio,boguliubov}.

The resulting correlated Green's functions  $G_c^{nm}$ have a structure with eight poles
\begin{equation}
    G_c^{(nm)}(\vek,\iom)=\sum_{j=1}^{8}\frac{Z^{(nm)}_{j}(\vek)}{\iom-E_{j}(\vek)},
\end{equation}
with $E_{j}(\vek)$ and $Z^{(nm)}_{j}(\vek)$ representing the correlated quasiparticle bands and spectral weights, respectively.
In terms of the correlated Green's function $G_c^{(12)}(\vek,\iom)$, the SC  order parameter defined in Eq. (\ref{Delta}) becomes:
\begin{equation}
\Delta_{\vec{k}}=-\sum_{\vec{k^{'}}}V_{\vec{k},\vec{k^{'}}}\frac{1}{\beta}\sum_{n}G_c^{(12)}(\vek,\iom),
\label{ffgg11}
\end{equation}
where $\beta=\frac{1}{k_BT}$, in which $T$ is the temperature and $k_B$ the Boltzmann constant. By considering the conditions established in Eq. (\ref{Vkk})  for $V_{\vec{k},\vec{k'}}$ and performing the summation over the Matsubara frequencies, we get:
\begin{align}
\Delta=\frac{V}{L}\sum_{\vec{k}}\sum_{j=1}^{8}Z^{(12)}_{j}(\vek)f\left(E_{j}(\vek)\right),
\label{gap}
\end{align}
where $f(E)$ is the Fermi function and $L$ is the number of lattice sites in the system.
Applying the process described above for the CDW order parameter, we have:
\begin{equation}
     W=-\frac{V}{L} \sum_{\vec{k}}\sum_{j=1}^{8}Z^{(13)}_{j}(\vek)f\left(E_{j}(\vek)\right),
     \label{W01}
\end{equation}
where $Z^{(13)}_{j}$ are the spectral weights of the correlated Green's function $G_c^{(13)}(\vek,\iom)$.

Finally, the total occupation is given by:
\begin{equation}
n_T=\frac{1}{L}\sum_{\vec{k}}\frac{1}{\beta}\sum_{n}[G_c^{(11)}(\vek,\iom)+G_c^{(33)}(\vek,\iom)].
\label{eqnT}
\end{equation}
Eqs. (\ref{gap})-(\ref{eqnT}) form a system of coupled equations that must be solved self-consistently.

\begin{figure}[!ht]
\begin{center}
 \includegraphics[angle=0,width=8.5cm]{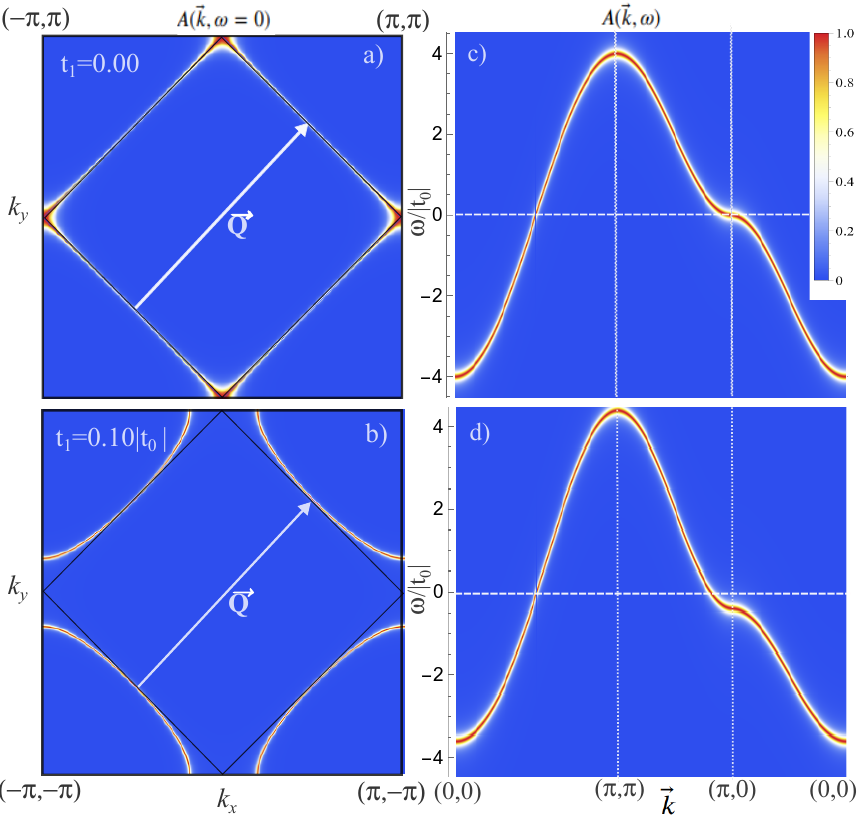}
\end{center}
\caption{ Panels a) and b) show the spectral function $A(\vec{k},\omega=0)$ in the normal state, for two different values of $t_1$. The black lines serve as a guide to the eye and indicate the locus of the perfectly nested Fermi surface.  Panels c) and d) show $A(\vec{k},\omega)$ along principal directions. The horizontal dashed line indicates the position of the Fermi energy $\varepsilon_F$. The temperature and the Coulomb interaction are $T=0$ and $U=0$, respectively.}
\label{fig:SF}
\end{figure}

\section{Numerical Results}

All numerical results presented in this section are obtained considering $t_0 = -1.0$ eV, with $|t_0|$  as the energy unit. For the nesting vector, we use the simplest case for a squared lattice, i. e., $\vec{Q} = (\pi,\pi)$. In addition, the chemical potential is fixed at $\mu = 0$ and the cutoff energies are $\varepsilon_{SC} = 0.2|t_0|$ and $\varepsilon_{CDW} = 1.1|t_0|$.

The retarded Green's function  $G^{11}_R(\vec{k},\omega)$ can be obtained from $G_c^{(11}(\vek,\iom)$ via analytical continuation \cite{mahan,gubernatis}. In terms of  $G^{11}_R(\vec{k},\omega)$, the spectral function is:
\begin{equation}
A(\vec{k},\omega)=-\frac{1}{\pi}\operatorname{Im} G^{11}_R(\vec{k},\omega).
\label{eqAwk}
\end{equation}
We can also introduce the density of states as $\rho(\omega)=\frac{1}{L}\sum_{\vec{k}}A(\vec{k},\omega)$.
The spectral function $A(\vec{k},\omega)$ for the normal state with two different values of $t_1$, is shown in Fig. \ref{fig:SF}. In Figs. \ref{fig:SF}a)  and \ref{fig:SF}b), the topology of the Fermi surface is displayed through the spectral function $A(\vec{k},\omega=0)$.
At half-filling and $t_1=0$, we observe (in panel \ref{fig:SF}a)) a perfect nesting for the chosen $\vec{Q}$ vector. In this case, the energy of the van Hove singularity in $(\pi,0)$ coincides with the Fermi energy $\varepsilon_F=0$, as can be seen in \ref{fig:SF}c). As a consequence, the CDW instability is favored even for small values of $V$. However, for $t_1=0.10|t_0|$, the energy of the van Hove singularity is shifted to lower energies, and only a portion of the Fermi surface presents perfect nesting, as shown in Fig. \ref{fig:SF}b).

\subsection{The CDW state}
This section presents results for the CDW state at zero temperature and in the absence of superconductivity.
\begin{figure}[!ht]
\begin{center}
\includegraphics[angle=0,width=6.5cm]{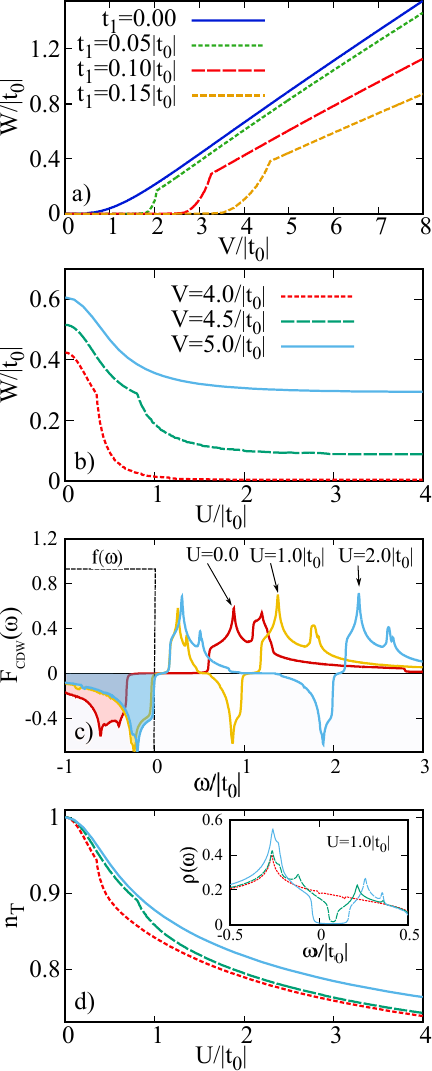}
\end{center}
\caption{\label{fig:POV} a) The CDW ($W$) order parameter as a function of the attractive interaction $V$, for $U=0$ and different values of $t_1$. b)
$W$ as a function of the Coulomb interaction $U$, for $t_1=0.10|t_0|$ and different values of $V$. c) Function $F_{CDW}(\omega)$ for $V=5.0|t_0|$, $t_1=0.10|t_0|$ and different values of $U$.
d) The total occupation $n_T$ versus $U$ for the same values of $V$ and $t_1$ as in b). The inset shows the density of states for $t_1=0.1|t_0|$.  The temperature is fixed at $T = 0$.}
\end{figure}
The effect of the second-nearest neighbor hopping amplitude $t_1$ on the CDW state is shown in Fig.  \ref{fig:POV}a), which exhibits $W$ as a function of $V$ for $U=0$ and different values of $t_1$.  The results reflect the fact that for $t_1=0$, the Fermi surface presents perfect nesting, therefore, a small value of $V$ is enough to stabilize the CDW state. On the other hand, for finite values of $t_1$, the Fermi surface is no longer perfectly nested, and  higher values of $V$ are necessary to stabilize the CDW state. This finding agrees with the previous result reported in reference \cite{balseiro}.  Fig. \ref{fig:POV}b) shows the effect of the Coulomb interaction $U$ on the order parameter $W$. In general, the increase of $U$ results in a decrease in $W$. For $V=4.0|t_0|$,  $W$ decreases until a value close to zero at $U\sim|t_0|$. However, for $V=4.5|t_0|$ and $5.0|t_0|$, $W$ initially decreases, but above a certain value of $U$, it reaches an approximately constant value. In order to better understand the effect of $U$ on $W$, we perform an analytical continuation \cite{mahan} on the Green's function $G_c^{(13)}(\vek,\iom)$ and rewrite the equation of the CDW  order parameter as:
\begin{equation}
     W=-V\int_{-\infty}^{\infty}f(\omega)F_{CDW}(\omega)d\omega,
     \label{W02}
\end{equation}
with
\begin{equation}
F_{CDW}(\omega)=\frac{1}{L}\sum_{\vec{k}}\sum_{j=1}^{8}Z^{(13)}_{j}(\vek)\delta(\omega- E_{j}(\vek)),
\label{Fcdw}
\end{equation}
where $\delta(x)$ is the Dirac delta.
Fig. \ref{fig:POV}c) shows the function $F_{CDW}(\omega)$ for different values of $U$. The CDW order parameter is directly related to the integral in $\omega$ of the product $f(\omega)F_{CDW}(\omega)$ (see Eq.(\ref{W02})).  At $T=0$, the Fermi function $f(\omega)$ is zero for $\omega >0$, thus only the range with $\omega \leq 0$ contributes to the value of $W$ \cite{calegari2005}. However, it is observed that in this range of $\omega$, the function $F_{CDW}(\omega)$ does not change significantly when $U\gtrsim|t_0|$, causing the value of $W$ to stabilize, as shown in Fig. \ref{fig:POV}b). Fig. \ref{fig:POV} d) shows the decrease in the total occupation $n_T$ due to the increase of $U$. As can be seen in the inset, for larger values of $V$, the opening of the CDW gap around $\omega =0$ moves states to the region below $\omega =0$, keeping  $n_T$ higher when compared with the cases with lower values of $V$.
%
\begin{figure}[!ht]
\begin{center}
\includegraphics[angle=0,width=8cm]{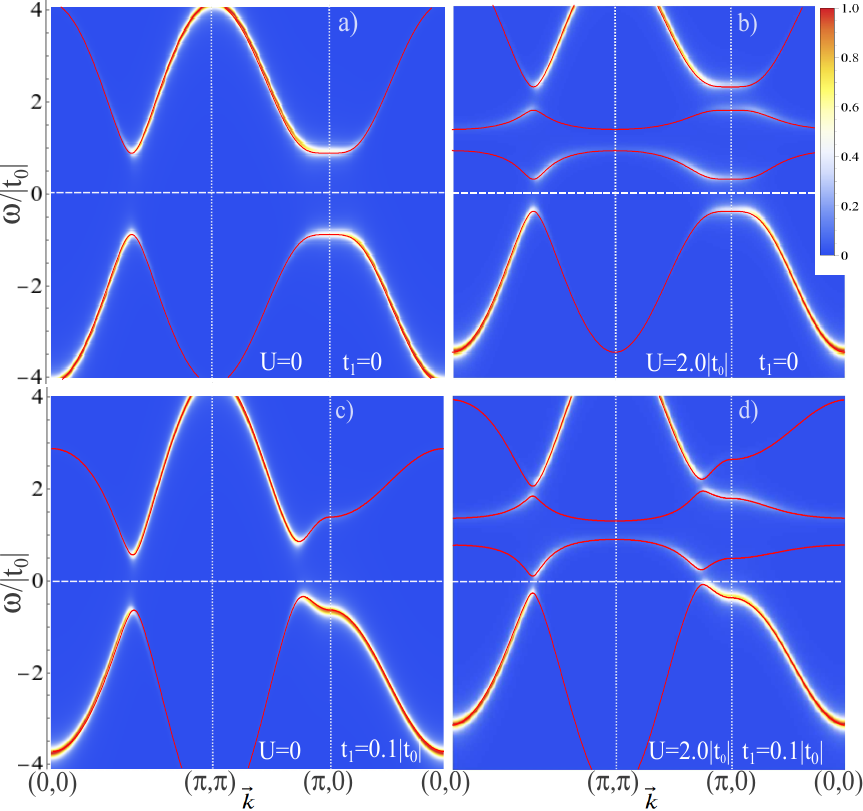}
\end{center}
\caption{\label{fig:bandasCDW} The spectral function $A(\omega,\vec{k})$ for $T = 0$ and $V=5.0|t_0|$. The color map indicates the intensity of the spectral weight. The solid red lines show the quasiparticle bands along principal directions. The horizontal dashed line, indicates the position of the Fermi energy $\varepsilon_F$.}
\end{figure}

Fig. \ref{fig:bandasCDW} shows the overlap of the spectral intensity $A(\omega,\vec{k})$ and the quasiparticle bands  for $V=5.0|t_0|$ and different values of $U$ and $t_1$. In Fig. \ref{fig:bandasCDW}a), with $U=t_1=0$, the CDW gap occurs symmetrically around $\omega=0$, both in the $(0,0)-(\pi,\pi)$ direction and in the region of the point $(\pi,0)$. In Fig \ref{fig:bandasCDW}b),  we have $U=2.0|t_0|$ while $t_1$ is kept null. In this case, we notice the presence of four quasiparticle bands and the
decrease of the CDW gap around $\omega=0$, due to the $U$ effect.
The effect of $t_1$ for the uncorrelated case is shown in Fig. \ref{fig:bandasCDW}c). In the region of point $(\pi,0)$,  an asymmetry of the CDW gap is observed relative to $\omega=0$. This asymmetry affects the nesting of the Fermi surface (see Fig. \ref{fig:SF}b)) and, as a consequence, it decreases the width of the CDW gap. Fig. \ref{fig:bandasCDW}d) shows  a significant suppression of the CDW gap due to the combined effects of both $U$ and $t_1$.

\subsection{Coexistence of CDW and SC}

\begin{figure}[!ht]
\begin{center}
\includegraphics[angle=0,width=8.5cm]{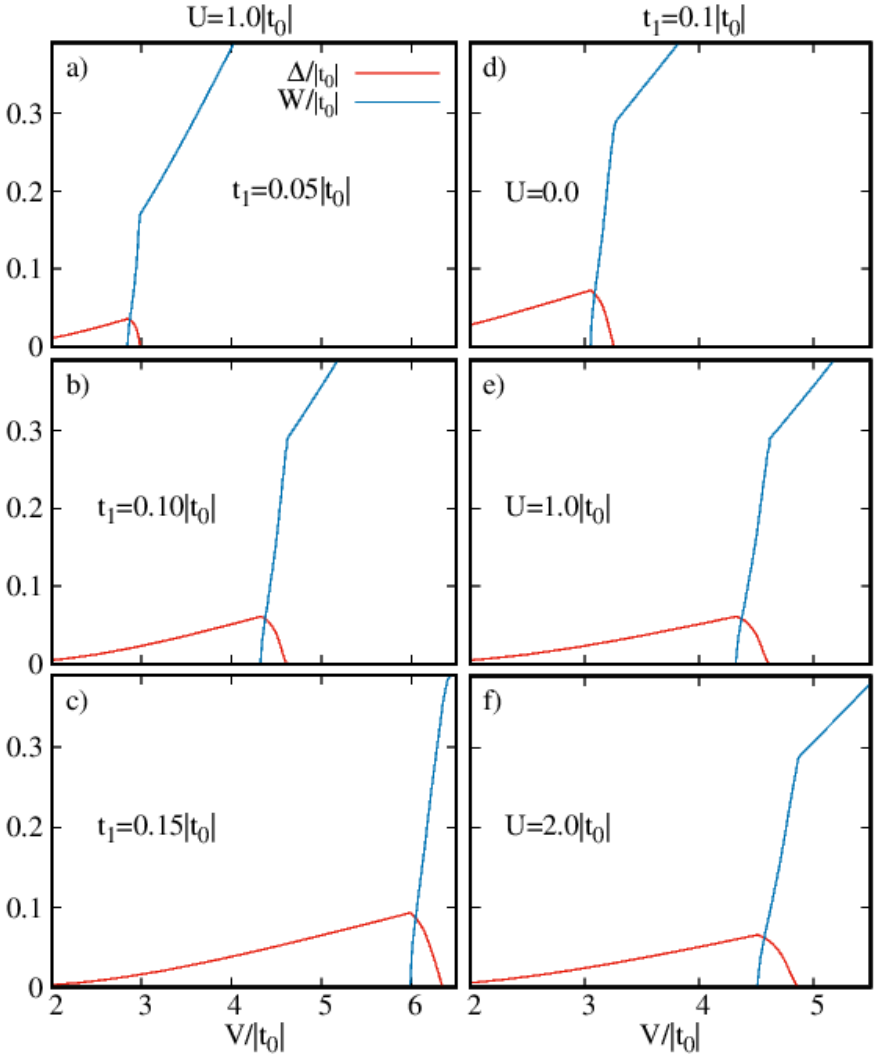}
\end{center}
\caption{\label{fig:POsV} The CDW ($W$) and SC ($\Delta$) order parameters as a function of the attractive interaction $V$, for different values of $U$ and $t_1$. The temperature is fixed at $T=0$.}
\end{figure}

Fig. \ref{fig:POsV} shows the behavior of the CDW and SC order parameters as functions of the attractive interaction $V$, at zero temperature.
At $V=2.0|t_0|$, in most cases the system is in the superconducting state, but if $V$  increases, the system evolves into a coexisting phase in which both $W$ and $\Delta$, are finite.  If $V$ continues to increase, $\Delta$ approaches  zero indicating that SC is suppressed and only the CDW state remains.

In particular, Figs. \ref{fig:POsV}a), \ref{fig:POsV}b) and \ref{fig:POsV}c), show the effect of the second-nearest neighbor hopping $t_1$ on the order parameters $W$ and $\Delta$, for $U=1.0|t_0|$. Due to the increase of $t_1$, a larger value of $V$ is required for the CDW to emerge. As observed in the case of a pure CDW,  the increase of $t_1$  reduces the nested regions of the Fermi surface; consequently,  a larger value of $V$ is needed to stabilize the CDW state. However, it can be noted that the coexistence region continues to exist, but it is shifted to higher values of $V$. The effect of the Coulomb interaction $U$  on the order parameters is shown in Figs. \ref{fig:POsV}d), \ref{fig:POsV}e) and \ref{fig:POsV}f), for $t_1=0.10|t_0|$. In general, the effect of $U$ on the order parameters is similar to the $t_1$ effect, i.e., $U$ shifts to higher values of $V$, the CDW state.
The present result for $U=0$ is in agreement with that  reported by Balseiro and Falikov \cite{balseiro}, who explored the interplay  between SC and CDW in the spirit of the BCS theory.

\begin{figure}[!ht]
\begin{center}
\includegraphics[angle=0,width=8.3cm]{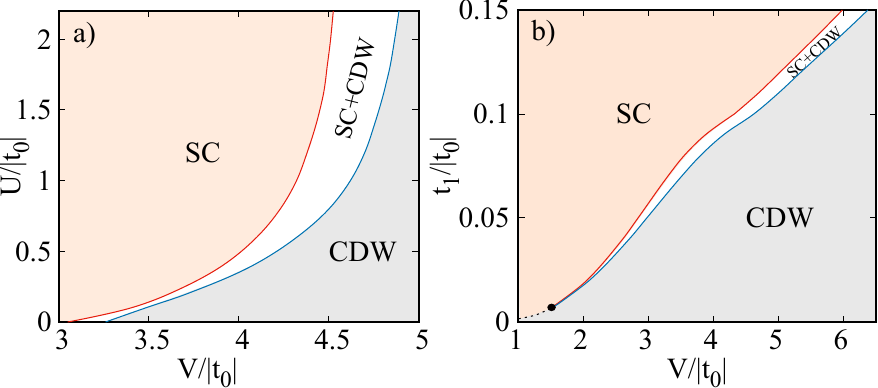}
\end{center}
\caption{\label{fig:DF} (a) Phase diagram for $U/|t_0|$ versus the attractive interaction $V$ with $t_1=0.10|t_0|$.
(b) Phase diagram for $t_1/|t_0|$ versus $V$ with $U=1.0|t_0|$. The temperature is fixed at $T=0$.}
\end{figure}

The effects of $U$ and $t_1$ on the superconducting and CDW states for $T=0$, are summarized in Fig. \ref{fig:DF}.
The phase diagram of $U$ versus $V$ is shown in Fig. \ref{fig:DF}a).  The region in which SC and  CDW coexist is shown in white.  Note that the coexistence region is present even for $U=0$. Furthermore, the range in $V$ for which the coexistence occurs, tends to remain almost constant at larger values of $U$. This behavior occurs due to the effect of $U$ on $W$ which leads to a stabilization of the $W$ value when $U$ exceeds $|t_0|$, as discussed in the case of a pure CDW state presented in Fig. \ref{fig:POV}c).
Fig. \ref{fig:DF}b) shows the phase diagram for $t_1$ versus $V$ with $U=1.0|t_0|$. The solid lines
represent second order transitions while the dashed line indicates an abrupt change in the order parameters
characterizing a first order transition between the SC and CDW states.
Considering that the increase of $t_1$ and pressure are closely related \cite{sil}, the present results agree, at least qualitatively, with a recent study with the pyrite-type superconductor CuS$_2$ showing that the increase of pressure gradually suppresses CDW and favors superconductivity \cite{shi}.

\begin{figure}[!ht]
\begin{center}
\includegraphics[angle=0,width=8.3cm]{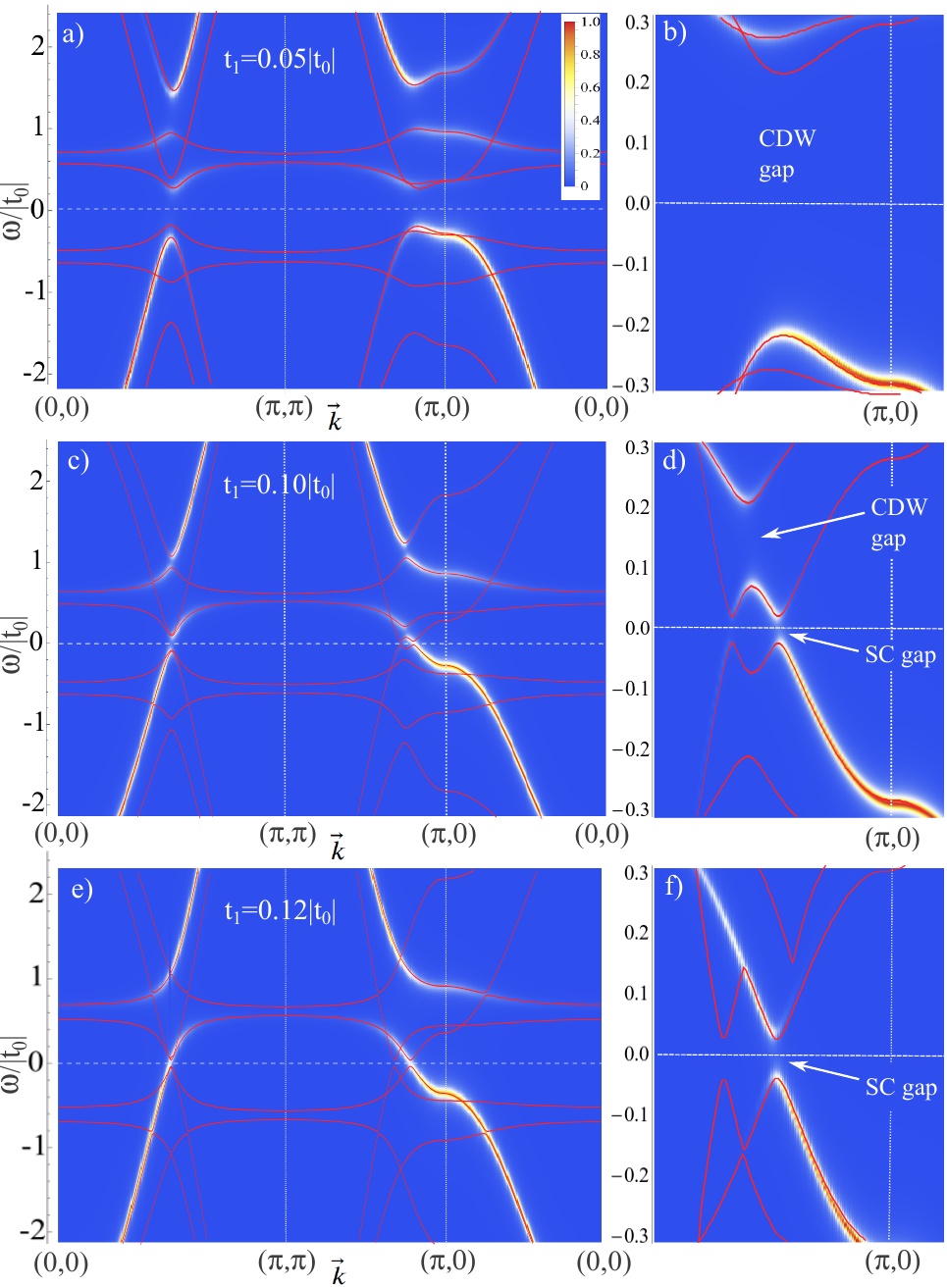}
\end{center}
\caption{\label{fig:bandascoex} The spectral function $A(\omega,\vec{k})$ and the quasiparticle bands (red solid lines) along principal directions for $V=4.5|t_0|$, $U=1.0|t_0|$ and different values of $t_1$. Panels a), c), and e) show an overview of the bands and spectral function for the respective $t_1$ values.
Panels b), d), and f) shown in detail the region near the antinodal point $(\pi,0)$, in order to highlight the presence of gaps for each case.  The horizontal dashed line, indicates the position of the Fermi energy $\varepsilon_F$ and the temperature is fixed at $T=0$.}
\end{figure}

In order to better understand the coexistence of the SC and CDW states, in  Fig. \ref{fig:bandascoex} we show the spectral function superimposed with the quasiparticle bands for $U=1.0|t_0|$, $V=4.5|t_0|$ and different values of $t_1$. Figs. \ref{fig:bandascoex}a) and \ref{fig:bandascoex}b) show the result for a small value of $t_1$, more precisely $t_1=0.05|t_0|$. Along the direction $(0,0)-(\pi,\pi)$ shown in Fig. \ref{fig:bandascoex}a), both the CDW and SC gaps, which are symmetric relative to the Fermi energy $\varepsilon_F=0$,  should occur simultaneously at the point $(\frac{\pi}{2},\frac{\pi}{2})$. However, in the direction $(\pi,\pi)-(\pi,0)$, the gaps compete to occur at the same energy and $\vec{k}$ ranges. But, since $t_1$ is small, the nesting of the Fermi surface is nearly perfect, therefore, the CDW state prevails over the SC state. As can be verified in Fig. \ref{fig:DF}b), for $t_1=0.05|t_0|$ and $V=4.5|t_0|$, the system is found in the CDW state. Fig. \ref{fig:bandascoex}b) shows in detail the CDW gap at the region of the point $(\pi,0)$. The analysis of the spectral function intensity allows us to identify the CDW gap at $-0.2\lesssim\frac{\omega}{|t_0|}\lesssim 2.8$. The result for $t_1=0.10|t_0|$ is shown in Figs. \ref{fig:bandascoex}c) and \ref{fig:bandascoex}d). In this case, the increase in $t_1$ from $t_1=0.05|t_0|$ to $t_1=0.10|t_0|$, is enough to shift the van Hove singularity to lower energies, causing a weakening of the CDW correlations and allowing the coexistence of the CDW and SC phases. Fig. \ref{fig:bandascoex}d) shows in detail the presence of the SC gap around the Fermi level $\varepsilon_F$, while the CDW gap occurs in an energy range above it. As discussed in the analysis of the pure CDW state, in the direction $(\pi,\pi)-(\pi,0)$, the presence of $t_1$ shifts the CDW gap to higher energies as shown in Figs. \ref{fig:bandasCDW}c) and  \ref{fig:bandasCDW}d). As a consequence, the SC and  CDW gaps emerge in different regions of the first Brillouin zone, reducing the competition between their states and even coexisting. If $t_1$ increases more, the nesting of the Fermi surface is completely destroyed and the SC phase prevails as can be seen in the Figs. \ref{fig:bandascoex}e) and \ref{fig:bandascoex}f), for $t_1=0.12|t_0|$. The results in Fig.  6 are consistent with the phase diagram shown in Fig. \ref{fig:DF}b).

\begin{figure}[!ht]
\begin{center}
\includegraphics[width=7.5cm]{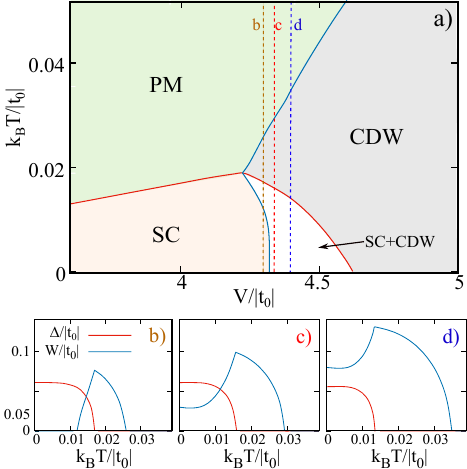}%
\end{center}
\caption{\label{fig:DFTvsV} a) Phase diagram for  temperature versus $V$ with $U=1.0|t_0|$ and $t_1 = 0.10|t_0|$. CDW ($W$) and SC ($\Delta$) order parameters as a function of $k_BT/|t_0|$ for $V=4.30|t_0|$ in b), $V=4.34|t_0|$ in c) and  $V=4.40|t_0|$ in d).}
\end{figure}

The effect of temperature $T$ on the SC and CDW states is shown in Fig. \ref{fig:DFTvsV}. The phase diagram of the temperature versus $V$ is shown in Fig. \ref{fig:DFTvsV}a) for $U=1.0|t_0|$ and $t_1=0.1|t_0|$.   The increase in temperature suppresses both the SC and CDW states, however the CDW phase that occurs in the region of most intense $V$ has a higher transition temperature. Moreover, the coexisting region decreases with the increase of $T$. Figs. \ref{fig:DFTvsV}b), \ref{fig:DFTvsV}c)
and \ref{fig:DFTvsV}d) shows the behavior of the order parameters for some particular values of $V$ in the coexistence region. The dashed vertical lines in \ref{fig:DFTvsV}a) indicate the values of $V$ for which the order parameters were calculated in Figs. \ref{fig:DFTvsV}b), \ref{fig:DFTvsV}c) and \ref{fig:DFTvsV}d). There is a direct correspondence between the colored letters that identify each figure and the colors of the lines.  In Figs. \ref{fig:DFTvsV}b), at low temperatures the system is found in the superconducting state characterized by $\Delta \neq 0$ and $W=0$. However, there is a region of $k_BT$ between $0.01$ and $0.02$, where both the order parameters are finite indicating that the system is in the coexistence regime. In Fig. \ref{fig:DFTvsV}c), the system is in the coexistence regime even at zero temperature, however, the amplitude of the superconducting order parameter is larger than that of the CDW order parameter. In Fig. \ref{fig:DFTvsV}d), the superconducting state is weaker and the amplitude of the CDW order parameter is larger even at zero temperature.  It is important to note that the main effect of $U$ is to suppress the CDW and SC phases, thereby decreasing the magnitudes of the order parameters without altering their qualitative temperature dependence. For this reason, the results shown in Figs.  \ref{fig:DFTvsV}b)-\ref{fig:DFTvsV}d) for the order parameters are in qualitative agreement with those reported in references \cite{Ekino}, \cite{Gabovich} and \cite{machida}, for models that do not explicitly incorporate the Coulomb interaction term.

\section{Conclusions}

In the present work we investigated the interplay between CDW and SC in the presence of Coulomb interaction $U$ and second-nearest neighbor hopping $t_1$.
The analysis of the spectral function in the half-filled normal state shows that the increase of $t_1$ partially destroys the perfect nesting of the Fermi surface. As a result, in the pure CDW state, the increase of $t_1$ produces an asymmetry of the CDW gap relative to the Fermi energy, especially near the point $(\pi,0)$. As a consequence, a larger value of $V$ is required to stabilize the CDW state.  The presence of the local repulsive Coulomb interaction $U$ splits the uncorrelated band $\varepsilon_{\vec{k}}$ causing the CDW gap to narrow on the quasiparticle bands. Because of this, the amplitude of the CDW order parameter $W$ decreases with $U$; however, it tends to stabilize,
above a given value of $U$. In general, both $t_1$ and $U$ are found to suppress the CDW state.

If, in addition to the CDW,  we also consider a superconducting  state, it is found that for certain sets of model parameters, the CDW and SC may coexist. It is important to note that, due to correlations related to $U$, larger values of the attractive interaction $V$ are required to enable this coexistence.  It has also been verified that  the presence of the second-nearest neighbor hopping $t_1$ is essential for coexistence. The analysis of the quasiparticle bands shows that, for small values of $t_1$, the CDW and SC gaps compete for the same region of the first Brillouin zone. However, as $t_1$  increases, the CDW gap shifts to a different region of $\vec{k}$ space, reducing competition and allowing both states to coexist. Considering that we assumed the CDW and SC gaps with $s$-wave symmetry, it is important to note that the gaps occur along both high symmetry directions, $(0,0)-(\pi,\pi)$ and  $(\pi,\pi)-(\pi,0)$. However, the coexistence of these phases is ultimately determined by the behavior of the quasiparticle bands near the antinodal point $(\pi,0)$.  This result show that second-nearest neighbor hopping $t_1$ is an important factor for tuning the SC and CDW coexistence.

The analysis of the order parameters as a function of temperature further clarifies the interplay between  SC and CDW states. For instance, in Fig. \ref{fig:DFTvsV}b), the  superconducting state is the most stable at low temperatures; however, as temperature increases, the thermal effects weaken SC, allowing the CDW state to emerge and coexist with SC
in a given range of temperature. At higher temperatures, the system evolves into a pure CDW state.  This sequence of phase transitions is observed experimentally in several systems \cite{morosan2,shi,kordyuk}.

In summary, this work presents important results that contribute
to a better understanding of the mechanisms behind the remarkable phenomenon of SC and CDW coexistence in correlated systems.

Data will be made available on request.

\section*{Acknowledgments}
This work was partially supported by Coordena\c{c}\~ao de Aper\-fei\c{c}oa\-mento de Pessoal de N\'{\i}vel Superior (CAPES); Conselho Nacional de Desenvolvimento Cient\'ifico e Tecno\-l\'o\-gico  (CNPq) and
Funda\c{c}\~ao de Amparo \`a Pesquisa do Estado do Rio Grande do Sul (FAPERGS). L. C. Prauchner and S. G. Magalhaes thank the CNPq (Conselho Nacional de Desenvolvimento Científico e Tecnológico), grant: 200778/2022-6. S. G. Magalhaes and E. J. Calegari also acknowledge the support of the INCT project Advanced Quantum Materials, involving the Brazilian agencies CNPq (Proc. 408766/2024-7), FAPESP, and CAPES.

\end{document}